# Itinerant electrons, local moments, and magnetic correlations in pnictides high temperature superconductors

P. Vilmercati[1], A. Fedorov[2], F. Bondino[3], F. Offi[4], G. Panaccione[5], P. Lacovig[6],
L. Simonelli[7], M. A. McGuire[8], A. S. M. Sefat[8], D. Mandrus[9], B. C. Sales[8],
T. Egami[1,8,9], W. Ku[10,11], and N. Mannella[1,*].

[1] Department of Physics and Astronomy, University of Tennessee, Knoxville, TN 37996, USA
[2] Advanced Light Source, Lawrence Berkeley National Laboratory, Berkeley, CA 94720, USA
[3] Laboratorio TASC, IOM-CNR, S.S. 14 km 163.5, Basovizza, I-34149 Trieste, Italy
[4] CNISM and Dipartimento di Fisica, Università Roma Tre, via della Vasca Navale 84, I-00146 Rome, Italy
[5] Istituto Officina dei Materiali (IOM)-CNR, Laboratorio TASC, Area Science Park, S.S.14, Km 163.5, I-34149 Trieste, Italy
[6] Sincrotrone Trieste S.C.p.A., Area Science Park, S.S. 14 Km 163.5, I-34149 Trieste, Italy
[7] European Synchrotron Radiation Facility, B.P. 220, F-38042 Grenoble, France
[8] Materials Science and Technology Division, Oak Ridge National Laboratory, Oak Ridge, TN 37831, USA
[9] Department of Materials and Engineering, University of Tennessee, Knoxville, TN, 37996, USA
[10] Condensed Matter Physics and Materials Science Department, Brookhaven National Laboratory, Upton, NY 11973, USA
[11] Physics Department, State University of New York, Stony Brook, NY 11790, USA
*nmannell@utk.edu

## ABSTRACT

A direct and element-specific measurement of the local Fe spin moment has been provided by analyzing the Fe 3$s$ core level photoemission spectra in the parent and optimally doped CeFeAsO$_{1-x}$F$_x$ (x = 0, 0.11) and Sr(Fe$_{1-x}$Co$_x$)$_2$As$_2$ (x = 0, 0.10) pnictides. The rapid time scales of the photoemission process allowed the detection of large local spin moments fluctuating on a $10^{-15}$ s time scale in the paramagnetic, anti-ferromagnetic and superconducting phases, indicative of the occurrence of ubiquitous strong Hund's magnetic correlations. The magnitude of the spin moment is found to vary significantly among different families, 1.3 $\mu_B$ in CeFeAsO and 2.1 $\mu_B$ in SrFe$_2$As$_2$. Surprisingly, the spin moment is found to decrease considerably in the optimally doped samples, 0.9 $\mu_B$ in CeFeAsO$_{0.89}$F$_{0.11}$ and 1.3 $\mu_B$ in Sr(Fe$_{0.9}$Co$_{0.1}$)$_2$As$_2$. The strong variation of the spin moment against doping and material type indicates that the spin moments and the motion of itinerant electrons are influenced reciprocally in a self-consistent fashion, reflecting the strong competition between the antiferromagnetic super-exchange interaction among the spin moments and the kinetic energy gain of the itinerant electrons in the presence of a strong Hund's coupling. By describing the evolution of the magnetic correlations concomitant with the appearance of superconductivity, these results constitute a fundamental step toward attaining a correct description of the microscopic mechanisms shaping the electronic properties in the pnictides, including magnetism and high temperature superconductivity.

One systematic, key aspect of almost all unconventional superconductivity, as observed in high-Tc cuprates and heavy fermions, is the resilient magnetic correlations in the superconducting state [1]. The same has been observed in the newly discovered Iron-based superconductors (Fe-SC), which offer the possibility of studying the relation between high-temperature superconductivity and magnetic correlations in a wide range of magnetic element-based materials [2].

Recent theoretical and experimental results suggest that the nature of the magnetic correlations in Fe-SC encompasses both the presence of itinerant electrons and local spin moments (LSM) [3,4,5,6,7,8,9,10,11,12,13,14]. While in chalcogenides there is agreement regarding the values of the LSM measured with different techniques and theoretical calculations, for the pnictides the situation remains puzzling [15]. For the pnictides 122, 111 and 1111 families, agreement regarding the magnitude of the LSM is lacking both between theory and experiments, and among different experiments as well. Specifically, the LSM in the paramagnetic phase of 122, 111 and 1111 measured with x-ray emission spectroscopy (XES) is found to be ≈ 1$\mu_B$, which is consistent with the ordered moments reported in the 122 by other techniques, but larger than that found in 111 and 1111 [15]. Interestingly, while in general Density Functional Theory underestimates the magnitude of the LSM, in the pnictides the opposite happens, with an estimated value ≈ $2\mu_B$ [2]. It has been pointed out how this disagreement originates from the occurrence of fast fluctuations of the LSM whose dynamic develops on timescales of the electron dynamic ($10^{-15}$ s) [8]. The timescale of these fluctuations is shorter than the response time of conventional magnetic probes such as Mössbauer, Nuclear Magnetic Resonance (NMR) and Muon Spin Rotation ($\mu$-SR) spectroscopy, which therefore provide a time-averaged value of the LSM. It is thus very important to carry out measurements with a fast probe in order to determine the true magnitude of the fluctuating LSM.

In this Letter, we present the measurements of the magnitude of the LSM in 122 and 1111 parent and optimally doped pnictides using core level photoelectron spectroscopy (PES). PES probes the electronic structure on timescales ≈ $10^{-16}$ - $10^{-15}$ s, much faster than the typical ≈ $10^{-8}$ s - $10^{-6}$ s timescales of Mössbauer, NMR and $\mu$-SR, and still 1-2 orders of magnitude faster than inelastic neutron scattering (INS). In addition, PES is sensitive to the single-site LSM, as opposed to the correlated moments measured by INS. Our data reveal unprecedented large LSM fluctuating $10^{-16}$ - $10^{-15}$ s timescales in the PM, anti-ferromagnetic (AFM) and superconducting (SC) phases, indicative of the occurrence of ubiquitous strong Hund's magnetic correlations. While almost insensitive to changes in temperature, the magnitude of the LSM is found to vary against material type and, more interestingly, against doping levels, the latter being a behavior neither predicted nor observed. This phenomenology is of utmost importance for clarifying the relation between high-temperature superconductivity and magnetic correlations.

Polycrystalline $CeFeAsO_{1-x}F_x$ (x = 0, 0.11) and $Sr(Fe_{1-x}Co_x)_2As_2$ (x = 0, 0.10) high quality single crystals have been grown and characterized as reported elsewhere [16,17,18]. Both doped samples are optimally doped with superconducting transition temperatures of 38 K for $CeFeAsO_{0.89}F_{0.11}$ and 14 K for $SrFe_{1.8}Co_{0.2}As_2$. Bulk-sensitive hard x-ray photoemission (HAXPES) measurements (hv = 7596 eV) were carried out on beamline ID16 at the ESRF Synchrotron Facility using the Volume Photoemission (VOLPE) spectrometer [19] in a pressure lower than $1.5 \times 10^{-9}$ Torr and total instrumental resolution ≈ 450 meV. Additional low energy PES measurements (hv = 216 eV) were carried out on Beamline 12.0.1 at the Advance Light Source with total instrumental resolution ≈ 50 meV. The samples have been measured after been fractured ($CeFeAsO_{1-x}F_x$) or cleaved ($Sr(Fe_{1-x}Co_x)_2As_2$) in-situ at temperatures between 15 K and 30 K.

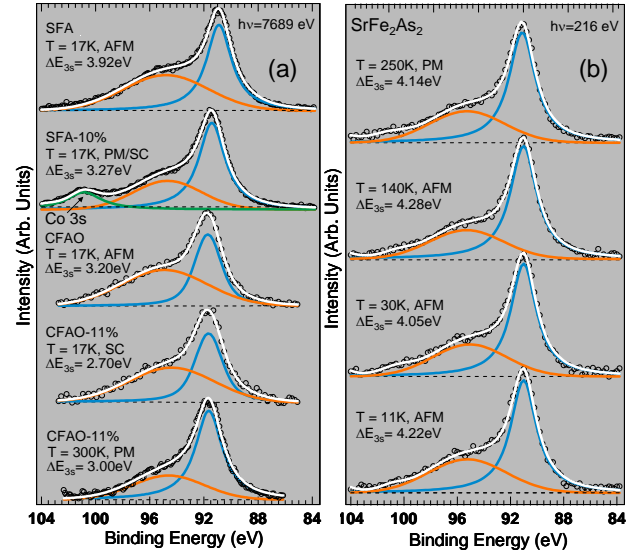

**Fig. 1. Multiplet splittings in Fe 3s core level HAXPES and PES spectra: evidence of strong on-site Hund coupling $J_H$ on Fe sites.** (a) HAXPES (hv = 7689 eV) Fe 3s core level spectra in CeFeAsO (CFAO), $CeFeAsO_{0.89}F_{0.11}$ (CFAO-11%), $SrFe_2As_2$ (SFA) and $Sr(Co_{0.12}Fe_{0.88})_2As_2$ (SFA-10%) at different temperatures in the antiferromagnetic (AFM), paramagnetic (PM) and superconducting (SC) phases. (b) PES (hv = 216 eV) Fe 3s core level spectra in $SrFe_2As_2$ at different temperatures in the PM and AFM phases. A Shirley-type background has been subtracted from the data points (circles). An additional background has been subtracted in the BE region ≈ 105 eV to account for some spectral weight originating from a nearby Auger peak for the spectra excited in $SrFe_2As_2$ with hv = 216 eV. The M-SP of the BE is clearly visible as a doublet structure consisting of a main line and a satellite peak at higher BE. The continuous white line through the data points is the result of the two component fit of the doublet (blue and orange lines). The distance between these two peaks maxima provides the multiplet separation $\Delta E_{3s}$, with experimental uncertainty on $\Delta E_{3s}$ of ± 100 meV in (a) and ± 50 meV in (b).

HAXPES and PES Fe 3$s$ core level spectra in different Fe-HTSC compounds are shown in Fig. 1. These spectra exhibit a doublet due to multiplet splitting (M-SP) of the binding energy (BE), a well-known effect in transition metals which provides a unique probe of the LSM of magnetic atoms [20,21]. The M-SP arises from the exchange coupling of the core 3$s$ electron left behind upon photoelectron emission with the net spin $S_V$ in the unfilled outer shell(s) of the emitter atom (Fe 3d/4s in this case) [20,21]. Fe 3$s$ photoelectrons have two different energies, and thus produce two different peaks, depending whether the spin of the electron left behind is parallel or antiparallel to $S_V$. The energy difference between the two peaks, referred to as multiplet energy separation $\Delta E_{3s}$, permits estimating the net spin $S_V$ of the emitter atom. To accomplish this, we follow a procedure which has been adopted for itinerant systems, and that has been proven to provide the correct value of $S_V$, and hence the LSM [22,23,24]. Specifically, work on metallic Mn and Co has shown that $\Delta E_{3s}$ scales linearly with $(2S_V + 1)$. The values for $S_V$ are obtained by extrapolating the linear fit of the measured splitting $\Delta E_{3s}$ plotted against $(2S_V + 1)$ for ionic compounds, for which $S_V$ is known since the valence is an integer number [22,23]. Multiplying the $S_V$ values by the spin factor $g = 2$, one correspondingly obtains the values for the Fe LSM. Importantly, the values of the LSM extracted from the 3s core-level spectra according to this procedure are remarkably close to the ones measured by Curie-Weiss type fits to magnetic susceptibility, ferromagnetic hysteresis loops, and neutron scattering studies of ordered states in metallic systems [22,23,24]. We follow this same approach given the itinerant character of the Fe-HTSC [24] (See Additional Information for further details).

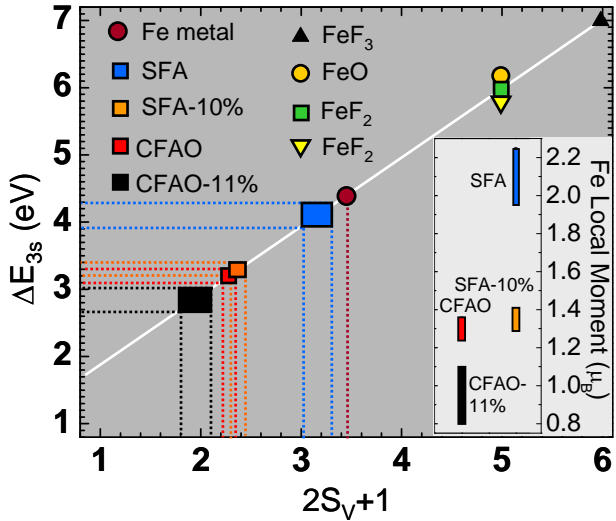

**Fig. 2. Estimate of the spin moment on the Fe sites from the multiplet energy separation $\Delta E_{3s}$.** The data points denote the values of the multiplet energy separation $\Delta E_{3s}$ for CeFeAsO (CFAO), CeFeAsO$_{0.89}$F$_{0.11}$ (CFAO-11%), SrFe$_2$As$_2$ (SFA) and Sr(Co$_{0.12}$Fe$_{0.88}$)$_2$As$_2$ (SFA-10%) at different temperatures and phases studied in this work. The continuous line is the extrapolation of the linear fit of the $\Delta E_{3s}$ values plotted against $(2S_V +1)$ for the Fe ionic compounds FeF$_3$, FeF$_2$, FeO, for which $S_v$ is known to be 5/2 (FeF$_3$) and 2 (FeF$_2$, FeO) [24]. The linear fit results in the relation $\Delta E_{3s} = 0.94 + 1.01 \times (2S_V+1)$. This procedure has been used to extract the values of $S_V$ from 3s core level spectra in itinerant systems such as metallic Mn, Co and Fe [24,22,23]. Notably, the value of the SM of metallic Fe is found to be $\approx 2.5$ $\mu_B$, remarkably close to the values of 2.2 and 2.33 $\mu_B$ measured with neutrons and magnetic susceptibility, giving us confidence in the correctness of this analysis procedure. The inset shows the magnitude of the SM. The size of the symbol is much bigger than the experimental uncertainties. It denotes the range of values found in this work for the splitting $\Delta E_{3s}$, the correspondent values for $S_V$, and the Fe local SM (inset).

The values of $\Delta E_{3s}$ are obtained with a two-component fit of the Fe 3s spectra, with uncertainty being estimated to be ± 100 meV and ± 50 meV for the HAXPES and PES spectra, respectively. The values of $\Delta E_{3s}$, the corresponding $S_V$ values, and the inferred values of the LSM obtained according to this procedure are shown in Fig. 2. We find that the values of the LSM are 1.3 $\mu_B$ in CeFeAsO and 2.1 $\mu_B$ in SrFe$_2$As$_2$, which decrease to 0.9 $\mu_B$ and 1.3 $\mu_B$ in the optimally doped samples, respectively. More specifically, the data show i) non vanishing LSM in all of the phases, ii) a non significant temperature dependence of the LSM, and iii) a marked dependence of the LSM upon doping. We now comment below on the implication of these observations.

Since M-SP occur exclusively in atoms with the outer subshell(s) partially occupied with a non vanishing net spin $S_V$, the Fe 3s spectra in Fig. 1 indicate that the electronic configuration on the Fe site is never found to be in the "low spin" state $S_v = 0$, indicating that LSM persist ubiquitously in different phases. LSM in the PM phase occurs either in the doped samples or above the Néel temperature $T_N$ for the parent compounds, in agreement with previous results [15,24]. LSM are also measured in the superconducting (SC) phase of CeFeAsO$_{0.89}$F$_{0.11}$, and in proximity of the SC/PM phase boundary in SrFe$_{1.8}$Co$_{0.2}$As$_2$. The presence of a LSM of similar magnitude has further been confirmed in the SC phase of SrFe$_{1.8}$Co$_{0.2}$As$_2$ with additional data not reported here. By exposing the occurrence of fluctuating LSM on the Fe sites with the same technique (PES), which has already revealed the itinerant character of the electrons [24], our findings provide unambiguous experimental evidence of the coexistence of LSM and itinerant electrons in the pnictides.

The large values of the Fe LSM indicate the occurrence of a rather strong on-site Hund coupling $J_H$ that fosters the electrons in the Fe $3d/4s$ shells to align parallel to each other, as suggested by theoretical investigations [3,7,25]. The values for the LSM shown here are significantly larger than the ordered moments detected in the AFM phase by Neutron Diffraction, Mössbauer spectroscopy, NMR, and $\mu$-SR [2]. Most notably, according to our measurements the LSM in SrFe$_2$As$_2$ amounts to 2.1 $\mu_B$, a value considerably higher than any experimental result reported so far, thus indicating a retrieval of the seemingly "missing" LSM in the 122 system [15]. The different time scales involved in the measurements can account for these marked differences. Since the photoemission process is fast ($\approx 10^{-16}$ - $10^{-15}$ s), the values of the LSM extracted from the analysis of the PES Fe 3s spectra are representative of the system sampled over extremely short time scales characteristic of electron dynamics (i.e. a snapshot). In contrast, the time scale of Mössbauer, NMR and $\mu$-SR measurements are typically $\approx 10^{-8}$ s - $10^{-6}$ s, practically static compared to the timescale of electrons dynamics. This discrepancy in the magnitude of the LSM between the fast ($\approx 10^{-16}$ s) and slow ($10^{-8}$ s - $10^{-6}$ s) measurements is due largely to the occurrence of quantum fluctuations, to which only fast measurements are sensitive. Considerations on electron dynamics provide a rationale for the signatures of both itinerant electrons and LSM exposed by the experiments. In the localized magnetism as found in insulating transition metal oxides and rare earth metals, LSM form from well localized electronic wavefunctions not participating in the Fermi surface (FS). In this case, the magnetism can be discussed concentrating on the magnetic degrees of freedom alone typically described by spin Hamiltonians, such as the Heisenberg Hamiltonian. This clean separation of magnetic and translational degrees of freedom does not occur in itinerant systems, since the magnetism stems from electrons which also happen to participate in the FS. A unique feature of itinerant systems, with no equivalence in localized magnetism, is that the amplitude of the LSM is not constant, but exhibit very fast fluctuations arising as a result of the electron dynamics. Itinerant electrons have wavefunctions which are phase-coherent over large distances, with the result that the electron density, and as a consequence the spin density, are not described by sharp quantum numbers. If $W$ denotes the bandwidth, itinerant systems are characterized by the presence of a fundamental timescale $\tau_F \approx h/W = \approx 10^{-15}$ s proper of electron motion [8]. On a timescale $\approx \tau_F$ the magnitude of the LSM is not constant since electrons cannot arrive at and leave a site with sufficient correlation between their spin orientations, thus setting the occurrence of very fast fluctuations in time referred to as *quantum fluctuations*. We stress that the quantum fluctuations are markedly different from what is commonly referred to as "spin fluctuations": The latter denote a slower wave-like precession of the atomic moments averaged over the fast quantum fluctuations, and correspond to the spin waves, whose time scales are $\tau_{SW} \approx h/W_{SW} \approx 10^{-14}$ - $10^{-13}$ s, much slower than the fast $\tau_F$.

Quantum fluctuations manifest directly in fast experiments with a short time constant $\approx \tau_F$, and thus involving large energy transfer. This is the case of the Fe 3s spectra in Fig. 1, whose analysis thus provides the values of the bare LSM $m_{loc}$, which corresponds to the response of the system on short time scales typical of fast quantum fluctuations [8]. Also the lineshape of the Fe 3s spectra is indicative of the occurrence of quantum fluctuations. First, the peak widths are intrinsically large, $\approx$ 2-3 eV, much larger than the experimental resolution. In addition, the best fits to the Fe 3s spectra are always obtained when the curve fitting the peak at higher BE- is mainly of Gaussian character, with a width much larger than that of the lower BE peak and than that expected from experimental resolution. Indeed, fluctuations in the amplitude of the LSM on Fe sites should appear in an Fe 3s spectrum as sidebands at higher BE with the peaks envelope being a Gaussian, reflecting the normal character of their distribution and the fact that $S_V$ is not a good quantum number [26]. On the contrary, conventional magnetic experiments average over fast quantum fluctuations since they probe the system on time scales much longer than $\tau_F$, with consequent low-energy transfer. They measure a screened moment which is strongly reduced as compared to the bare LSM $m_{loc}$ [8]. Although dynamical information can be obtained in INS experiments from integrating the spin susceptibility over energy and momentum [14], we stress that this analysis provides information of the *correlated* LSM, i.e. $\Sigma_j < S_j S_{i+j} >$, which is a different entity than the bare LSM $<S_i>$ measured in PES. The coexistence of local LSM and itinerant electrons indicates that the physics of Fe-SC is controlled essentially by different energy scales that correspond to different time limit of the dynamical response of the system: as extreme, one has a large ($\approx$ eV) energy scale, indicative of the quantum fluctuations, and a small ($\approx$ 1-100 meV) energy scale, which corresponds to dressed interactions. The large and small energy scales manifest in the magnetic response of the system as a bare LSM $m_{loc}$, and a screened LSM, respectively [8].

It is insightful to compare our findings with the results of XES measurements reported in [15]. On the one hand, our results are in agreement with those in [15] regarding the magnitude ($\approx 1.3~\mu_B$) of the SM for the parent 1111 and the doped 122 systems, and the overall insensitivity of the moment on temperature. On the other hand, our results show a large LSM $\approx 2.1~\mu_B$ in the SrFe$_2$As$_2$ system, and a significant reduction of the SM upon doping for both the 1111 and the 122 systems. Given that XES is a fast probe, especially for transition involving the *Fe K* shell, it is hard to reconcile these differences on the basis of different time scales involved in the measurements. It has been proposed that itinerant electrons are not properly counted in the XES detection of the Fe SM due to the local nature of the Fe 1s core-hole potential [15]. In the Fe 3s core levels PES spectra, the contribution of *Fe 3d* and *4s* electrons to the M-SP is weighted by the *effective* exchange integrals $J^{eff}_{3s\text{-}3d}$ and $J^{eff}_{3s\text{-}4s}$, and hence by the overlap of the 3s core level and the 3d/4s valence electrons wavefunctions. The larger overlap of the 3s core level with the valence electron wavefunctions as compared to the 1s core level provides an explanation as to why PES experiments are able to detect more sensitively valence electrons contributing to the SM than the XES measurements.

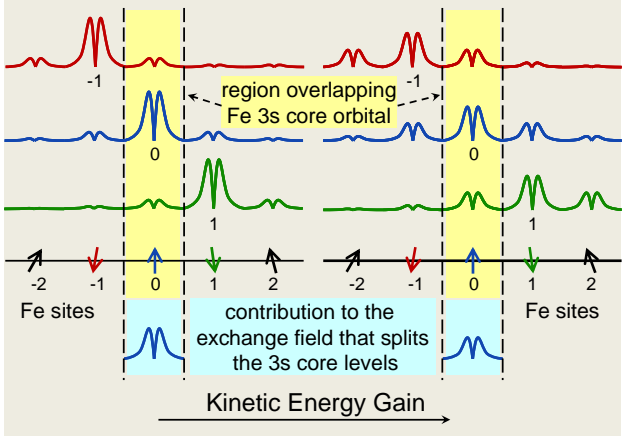

**Fig. 3. Reduction of the local moment upon enhanced kinetic energy gain.** Schematic of spatial distribution of the fluctuating spins centered at Fe sites (labeled by integer numbers), upon "integrating out" the higher-energy degree of freedoms that correspond to fluctuations faster than the PES time scale ($\approx 10^{-16}$ s). More precisely, the schematic illustrates the spin polarized Wannier orbitals that span the remaining low-energy fermionic space. With enhanced effective kinetic processes (upon doping away from integer number of electrons per Fe, applying pressure, or changing Fe-As-Fe bond angle), the spatial extent of the fluctuating spin increases as shown in the right panel, and consequently the central contribution decreases. In turn, the exchange field applied to the small region spanned by the 3s core orbital (within dashed lines) reduces, producing a smaller multiplet splitting in the 3s PES. Such a high-energy kinetic driven reduction is most visible in the presence of strong short-range AFM correlations, in which case the tails from neighboring spins gives opposite contributions.

A significant reduction of the LSM is found comparing the 122 parent compound with the 1111 parent compound. Even most notably, the reduction of the measured LSM is substantial upon doping in both families. This phenomenology is not compatible with a local-only nature of the LSM, as the local properties of the Fe ion against doping or materials type cannot change as much to justify the $\approx$ 40% reduction of the LSM. On the contrary, these observations reveal the important role played by the itinerant electrons in mediating the magnetism of the pnictides via interaction with the LSM. Interestingly, Hubbard and Hasegawa were among the first to propose an amalgam of localized and itinerant models when studying the magnetism in metallic Fe [27,28,29]. They pointed out that the motion of the itinerant electrons and the configurations of the exchange fields, entities essentially proportional to the local LSM of atoms, are influenced reciprocally in a self-consistent fashion [27,28,29]. In a context specific to the pnictides, it has been discussed how the interaction between the LSMs is mediated by the itinerant electrons in a self consistent fashion thanks to the provision of additional degrees of freedom such as the low electron kinetic energy and a two-fold orbital freedom, i.e. the degeneracy of the $d_{xz}$ and $d_{yz}$ orbitals [7,30]. The large contribution of the itinerant electrons in increasing the kinetic energy gain and the two-fold orbital degeneracy provide new degrees of freedom which add significant flexibility to how the itinerant electrons can interact with different local magnetic correlations.

The reduction of the measured LSM against doping and material type can be rationalized as a consequence of increasing the kinetic energy gain, achieved by spreading out the spatial distribution of the fluctuating spins (spin-polarized Wannier orbitals) onto multiple atomic sites (cf. Fig. 3). As a consequence, the exchange field at a particular site in the region spanned by the 3s core orbital is reduced, which in turn is responsible for the systematic reduction of $\Delta E_{3s}$ in the Fe 3s core level spectra, and hence the measured LSM. An additional effect responsible for the reduction of the exchange field at a particular site is the strong short-range AFM correlations, in which case the tails from neighboring spins give opposite contributions (cf. Fig. 3). The reduction associated with the large values of the LSM exposed by our data reflect the strong competition between the AFM super-exchange interaction among the LSM, and the kinetic energy gain of the itinerant electrons in the presence of a strong Hund's coupling [7]. Note that since the LSM fluctuate at a high frequency, the reported large LSM have irrelevant effects on the low-energy pair-breaking processes.

In conclusion, we presented experimental evidence of unprecedented large LSM fluctuating on quantum timescales in the PM, AFM and SC phase of pnictides using core level PES, an experiment sensitive to the single-site moment that probes the electronic structure on a much faster time scale than that of conventional magnetic probes. The data reveal a large LSM fluctuating on a $10^{-15}$ s timescale amounting to 2.1 $\mu_B$ in SrFe$_2$As$_2$ and 1.3 $\mu_B$ in CeFeAsO, that decreases to 1.35 $\mu_B$ and 0.9 $\mu_B$ in the optimally doped samples. The very large size of the LSM is evidence for the occurrence of strong Hund's magnetic correlations. The strong variation of the LSM against doping and material type indicates that the LSM and the motion of itinerant electrons are influenced reciprocally in a self-consistent fashion, reflecting the strong competition between the AFM super-exchange interaction among the LSM, and the kinetic energy gain of the itinerant electrons in the presence of a strong Hund's coupling. Our study encourages strongly development of future understandings of magnetism and superconductivity along similar lines of consideration, namely correlated metals under the influence of strong coupling to LSM.

**Acknowledgments**

The work at ALS and European Synchrotron Radiation Facility is supported by NSF grant DMR-0804902. The work at Oak Ridge is sponsored by the Division of Materials Science and Engineering, Office of Basic Energy Sciences. Oak Ridge National Laboratory is managed by UT-Battelle, LLC, for the U.S. Department of Energy under Contract No. DE-AC05-00OR22725. Portions of this research performed by Eugene P. Wigner Fellows at ORNL.

# SUPPLEMENTARY MATERIAL

## 1. Multiplet Splitting Effects in Fe 3s Core Level Photoemission Spectra.

Fig. 1 in the manuscript shows HAXPES and PES Fe 3s core level spectra in different Fe-SC compounds. These spectra exhibit a doublet due to multiplet splitting of the binding energy, a well-known effect in transition metals indicative of magnetic correlations. Multiplet splittings arise from the various possible non-degenerate total electronic states that can occur in the final states of the PES process. Such effect can occur only in systems in which the outer subshell(s) are partially occupied with a non vanishing spin $S_V$. The multiplet splitting arises from the coupling of the core electron left behind upon photoelectron emission with the net spin $S_V$ in the unfilled 3d/4s shells of the emitter atom, Fe in this case (cf. Fig. SM1). The analysis of multiplet splitting effects is considerably simplified for 3s core level spectra. In this case, since the core hole has zero angular momentum, the number of possible final states is considerably reduced [1,2]. For these reasons, PES 3s core level spectra have been employed as a probe of the local spin moment of magnetic atoms [1,2,3,4,5].

The interpretation of multiplet exchange splittings in terms of local spin moment is usually adopted for ionic systems [1-5]. According to a result known as Van Vleck's Theorem, the multiplet energy separation $\Delta E_{3s}$ depends on the net spin $S_V$ of the emitter via $\Delta E_{3s} = (2S_v + 1)J^{eff}_{3s-3d/4s}$, where $J^{eff}_{3s-3d/4s}$ denotes the effective exchange integral between the 3s and the 3d/4s shells after allowing for final-state intra-shell correlation effects [1,2]. This simple exchange-splitting interpretation of the 3s core level spectra is no longer complete as the number of electrons increases (as in the case of Cu and Ni oxides) and/or as the electronegativity of the ligand decreases, in which case charge transfer final-state screening effects become important and can lead to additional spectral structures [3,4,5]. For the vast majority of cases, the theoretical description of 3s core level spectra in transition metal oxides has been carried out with the so called Charge Transfer Multiplet (CTM) theory, in which both charge transfer and multiplet effects are taken into account. The essence of this approach is to describe the transition metal atom as surrounded by a cluster of ligands and consider the multiplet coupling effects within different electronic configurations due to charge transfer effects, i.e. charge fluctuations in the 3d states arising from hybridization effects between the TM atom and the ligand orbitals. The initial state is described as a linear combination of electronic configurations such as $d^n$, $d^{n+1}\underline{L}$, $d^{n+2}\underline{L}^2$ and alike configurations, where $\underline{L}$ denotes a hole in the ligand. The 3s core level spectra in Fe-based materials are always found to be largely dominated by exchange-splitting components, with no considerable charge transfer effects [3,4,5].

Although the analysis of 3s core levels has usually been carried out for ionic compounds, multiplet exchange splittings are also detectable in metallic systems [6,7,8]. Work on metallic Mn and Co has shown that the complications related to charge transfer effects describe above do not occur in metallic systems: Although Van Vleck's Theorem is insufficient to properly describe the itinerant nature of the electrons [6,8], $\Delta E_{3s}$ is found to scale linearly with $(2S_V + 1)$, indicating that the *3s-3d/4s* exchange interaction is the dominant contribution of the lineshape of the 3s core level spectra [6,8]. A description of the 3s core level spectra based on CTM Theory is very inadequate for systems of delocalized character such as metals. For metallic systems, the procedure used to extract the value $S_V$ from $\Delta E_{3s}$ consists in extrapolating the linear fit of the measured splitting $\Delta E_{3s}$ plotted against $(2S_V + 1)$ for compounds of unambiguous ionic character, for which $S_V$ is known since the valence is an integer number, as in the fluoride compounds [6,8]. Multiplying the $S_V$ values by the spin factor $g = 2$, one correspondingly obtains the values for the local spin moment. Importantly, the values of the local spin moment extracted from the 3s core-level spectra according to this procedure are remarkably close to the ones measured by Curie-Weiss type fits to magnetic susceptibility, ferromagnetic hysteresis loops, and neutron scattering studies of ordered states for Mn and Co [6,8]. More recently the same has been observed for metallic Fe: Notably, using the same approach, the value of the spin moment of metal Fe is found to be ≈ 2.5 μB, remarkably close to the values of 2.2 and 2.33 μB measured with neutrons and magnetic susceptibility, respectively, giving us confidence in the correctness of this analysis procedure [ 9 ]. We follow this same approach given the itinerant character of pnictide Fe-SC.

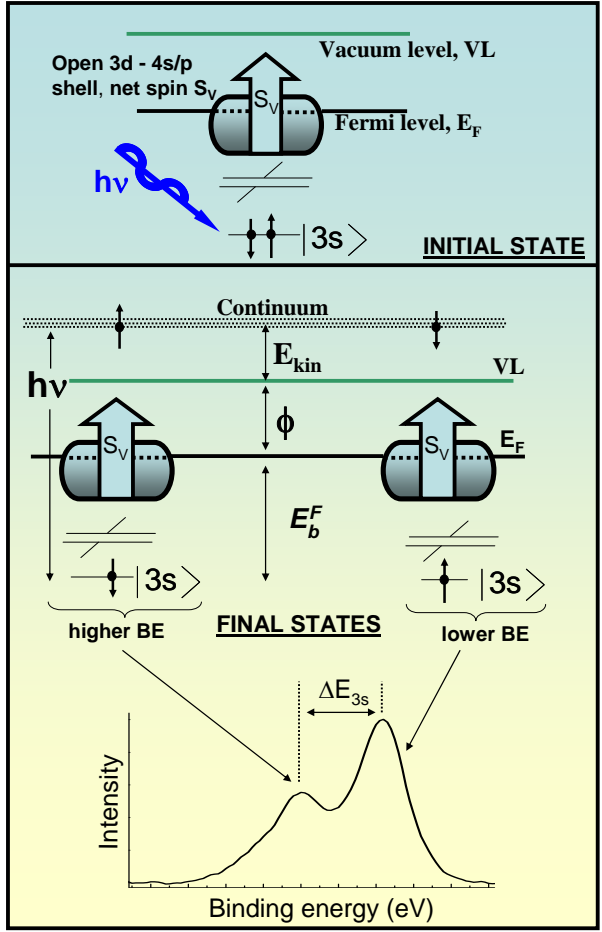

Fig. SM1. Multiplet splittings in Fe 3s core level Photoemission Spectra  Schematic layout of the multiplet splitting of the Binding Energy (BE) in 3s core level spectra of transition metals (TM). The upper panel shows the schematic energy levels of a TM atom with an unfilled shell with total net spin $S_V$ formed by electrons in the *TM 3d* and *4s/p* levels. The TM *3s* core levels host two electrons with opposite spins. Upon absorption of a photon of energy $h\nu$, electrons in the *3s* core levels are excited in the continuum above the vacuum level. For a system of N particles with ground state energy equal to $E_0(N)$, energy conservation in the photoemission process requires that $h\nu + E^0(N) = E_{KIN} + \phi + E^*(N-1)$, where $E_{KIN}$, $\phi$ and $E^*(N-1)$ denote the kinetic energy of the photoelectron, the work function, and the energy of the N-1-particle system in the presence of the core hole left behind upon photoelectron emission. Importantly, the asterisk (*) indicates that the final state of the photoemission process involves in general excited states of the remaining *N-1*-particle system. By detecting photoelectrons of kinetic energy $E_{KIN}$, the photoemission process allows the measurement of the BE defined as $E_B^F = h\nu - E_{KIN} - \phi = E^*(N-1) - E^0(N)$, where the *F* superscript indicates that the BE is referred to the Fermi level. This expression makes clear that final states $E^*(N-1)$ of lower (larger) energies are detected at lower (larger) BE. Upon emitting an electron from the *3s* core level, two final states are possible, corresponding to the configurations in which the remaining core *3s* electron is either parallel or anti-parallel to the net spin $S_V$ in the unfilled *3d-4s/p* shell of the TM atom. The exchange energy of the state with parallel spins is lower than that with anti-parallel spins.  The multiplet separation $\Delta E_{3s}$ corresponds to the energy difference between these two final states.

It is proper and important to discuss the results of a study of Fe 3s core level photoemission spectra recorded in several Fe crystalline and amorphous alloys more than two decades ago by van Acker et al. [10]. It is therein concluded that the Fe 3s splitting is not a reliable guide to local spin moments of Fe. This conclusion is based on the following observations: i) The ratio of the satellite to main peak intensity, commonly referred to as Branching Ratio, is always less than expected for either the atomic limit, or for itinerant magnetism, and ii) It is found that there is in general poor correlation between the Fe 3s splitting and the magnetic moment on the Fe atoms as measured with conventional magnetic techniques, with even Fe 3s splitting in some Pauli paramagnets.  We now comment on these two conclusions in order.

Caution should be in general advertized in using the branching ratio as a way to extract the value of $S_V$ for two different reasons.  First, the use of different photon energies results in a different branching ratio, which is smaller the lower the photon energy.  This behavior originates as a consequence of a change from an adiabatic to a sudden photoemission regime as the kinetic energy of the photoelectrons increases, as studied previously in both gas-phase and solid-state spectra of Mn [ 11 ].  Second, different experimental conditions such as different photon energies, detection directions and angular acceptance of the electron analyzer may result in different photoelectron diffraction effects which can affect the relative intensity of the two peaks.  In fact, in this work we have made no attempt in extracting the magnitude of $S_V$ from the branching ratio.

Van Acker et al. pointed out that there is in general poor correlation between the Fe 3s splitting and the magnetic moment on the Fe atoms.  In particular, Fe 3s splittings are measured in some Pauli paramagnets, which have no spin moment. These conclusions on the basis of the available body of data, are specious. In short, the argument proposed by Van Acker et al. ignores dynamical effects intimately connected with the existence of quantum fluctuations. The issue is clarifying what is meant by local spin moment. In ref. [10], the "spin moment" is obtained from conventional techniques such as Neutron Diffraction, Mössbauer spectroscopy, NMR, and $\mu$-SR which, by probing the system on timescales $\approx 10^{-8}$ s - $10^{-6}$ s, are practically static compared to the timescale of electrons dynamics. On the contrary, PES is a fast probe, with timescales $\approx 10^{-16}$ s - $10^{-15}$ s, and hence sensitive to fast quantum fluctuations which cannot be captured by slow techniques [12,13,14]. It is therefore possible that spin moments fluctuate on fast times typical of electron dynamics, but then are averaged to zero when measured with slow probes such as conventional magnetic techniques [12,13,14]. Remarkably, Pauli paramagnets such as $NbFe_2$, in which multiplet splittings can be observed but with no permanent spin moment as measured with conventional techniques, are materials exhibiting quantum critical point-behavior [15,16]. A similar occurrence is found in the intermetallic compound FeAl: this compound is known to be nonmagnetic with fluctuating magnetism, and the Fe 3$s$ core-level spectrum exhibits a two-peak structure [17,18]. In particular, FeAl is a material that is thought to be near a magnetic quantum critical point where it is subject to strong fluctuations in spin. We stress that when materials are nonmagnetic with an almost temperature-independent magnetic susceptibility (e.g., Rh metal), no multiplet splitting is measured in the 3$s$ spectrum. The lineshape of Fe 3$s$ spectra are indicative the occurrence of fast fluctuations in the magnitude of the spin moment. As discussed in our manuscript, the best fits to the Fe 3$s$ spectra in pnictides are always obtained when the curve fitting the peak at higher BE is mainly of Gaussian character, with a width much larger than that of the lower BE peak and that expected from experimental resolution. Indeed, fluctuations in the magnitude of the moment on Fe sites should appear in an Fe 3$s$ spectrum as sidebands at higher binding energies with the envelope of the peaks being a Gaussian, reflecting the normal character of their distribution. We stress that these fluctuations of the magnitude of the spin moments are profoundly different from spin fluctuations associated with local moment magnetism, i.e. spin moments with constant amplitude but with fluctuations in directions, as found for example in the high temperature paramagnetic phases of ionic compounds. In this case, the Fe 3s spectra exhibit two peaks with similar widths and correspond to a well difined value of the spin $S_V$, as all of the sites host the same value of the spin moment. In conclusion, a mismatch between the values of spin moments measured with different techniques is indicative of the occurrence of different physics with proper timescales.